\documentclass[twocolumn,final,10pt]{article}
\usepackage{times}
\usepackage{cite}

\makeatletter
\renewcommand\section{\@startsection {section}{1}{\z@}%
                                   {-3.5ex \@plus -1ex \@minus -.2ex}%
                                   {2.3ex \@plus.2ex}%
                                   {\sffamily\large\bfseries}}
\renewcommand\subsection{\@startsection{subsection}{2}{\z@}%
                                     {-3.25ex\@plus -1ex \@minus -.2ex}%
                                     {1.5ex \@plus .2ex}%
                                     {\sffamily\normalsize\bfseries}}
\renewcommand\subsubsection{\@startsection{subsubsection}{3}{\z@}%
                                     {-3.25ex\@plus -1ex \@minus -.2ex}%
                                     {1.5ex \@plus .2ex}%
                                     {\sffamily\normalsize\itshape}}
\makeatother

\title{\sffamily\bfseries{Personal Information Ecosystems and Implications for Design}}

\author{
  Manas Tungare, Pardha S. Pyla, Manuel P\'erez-Qui\~nones, and Steve Harrison 
  \\
  Dept. of Computer Science, Virginia Tech
  \\
  \{manas, ppyla, perez, srh\}@vt.edu
}

\date{}

\begin{document}
\maketitle

\abstract{
Today, people use multiple devices to fulfill their information needs. However, designers design each device individually, without accounting for the other devices that users may also use. In many cases, the applications on all these devices are designed to be functional replicates of each other. We argue that this results in an over-reliance on data synchronization across devices, version control nightmares, and increased burden of file management. In this paper, we present the idea of a \textit{personal information ecosystem}, an analogy to biological ecosystems, which allows us to discuss the inter-relationships among these devices to fulfill the information needs of the user. There is a need for designers to design devices as part of a complete ecosystem, not as independent devices that simply share data replicated across them. To help us understand this domain and to facilitate the dialogue and study of such systems, we present the terminology, classifications of the interdependencies among different devices, and resulting implications for design.
}

\section{Introduction}
\label{section-introduction}

The last few years have seen a massive proliferation of a variety of computing devices spread across a broad spectrum of capabilities and form factors. Each class of these devices has features and affordances that makes it unique from the others.  Users, however, use multiple devices in concert with one another, to accomplish their everyday computing tasks. Weiser's vision \cite{weiser_1991_the-computer} of embedding invisible computation into the environment is now a technological reality. However, the design of today's user interfaces is still done on a device-by-device basis.  There appears to be a need to design interfaces more globally to  support the spanning of a user's tasks across these multiple devices.

Today's dominant design trend is to think of mobile applications as clones of desktop versions that run on multiple platforms. Often, maintaining consistency across platforms has been the prime focus of design. For example, Microsoft Windows Mobile (for mobile devices) is a scaled-down version of desktop versions of Microsoft Windows, with a similar \textit{start} button and user interface widgets.  Similarly, calendar, address book, and email programs have been ported from the desktop platform to Personal Digital Assistants (PDAs) with only cosmetic changes. Most of these applications provide a duplication of functionality across devices because of what appears to be the designer's implicit assumption that a user would perform the same tasks equally on all devices. Even at a user interaction level, each device is often treated in isolation from other devices. For example, a user can set an alarm on their calendar software, and through synchronization, this alarm is often duplicated to a number of devices (e.g. laptop computer, PDAs, cellphones, and iPods). Since none of these devices is aware of the others, the inevitable outcome is the (almost) simultaneous ringing of all alarms at the appointed hour. Even more frustrating is that the user has to, at times, turn off the alarms individually. This demonstrates a lack of consideration on the designer's part for the users' use of these devices as a synergistic whole.

This approach of replicating similar functionality across platforms demands an explicit provision for each device to be able to synchronize data with another. The need for such explicit synchronization mechanisms highlights the fact that these devices were designed as disparate islands of information that need to be bridged together to be used effectively. The synchronization software and an explicit synchronization procedure is a requirement for a new device to be integrated into a user's work environment. This burdens users by requiring  them to engage in \textit{planful opportunism} \cite{perry_2001_dealing}. It also taxes them with extra efforts in keeping track of what files to copy and knowing which platform has the latest versions of their personal information. Further, this situation aggravates the existing problem of information overload and fragmentation \cite{boardman_2004_stuff, bergman_2006_the-project}.

We argue that part of this problem is that the HCI community lacks the appropriate terminology, concepts, and principles with which to study this problem of multiple devices. In this paper, we define this collection of devices as an ecosystem and draw parallels with biological systems as a way to understand the devices that participate in the ecosystem, the relationships among them, the type of user activities that the ecosystem supports, and the equilibrium that must be maintained. We argue that to design ecosystems of multiple devices, designers must take the responsibility of maintaining the ecosystem equilibrium when new devices are introduced.

\section{Ecosystems}

A biological ecosystem consists of organisms, the environment in which they reside, the interactions that these organisms have among themselves and with their environment, and the natural balance that must be maintained to keep the ecosystem in equilibrium. In the following section, we define a \textit{``personal information ecosystem''} and draw parallels between a user's information domain and biological ecosystems.

\subsection{Definition of a Personal Information Ecosystem}

A `personal information ecosystem'  can be defined as \textit{`a system of devices and applications that are present in the information environment of a user, that interact closely and richly with one another, to help the user achieve the goal of fulfilling his/her information needs.'} 
The following sections discuss the parallels between the two ecosystems in detail.

In order to illustrate our idea of an ecosystem, consider the example of the popular iPod~+~iTunes \footnote{http://www.apple.com/ipod/} multi-platform interface. The iTunes application supports media management, creation and editing of playlists, media playback, and other tasks through a desktop-sized multi-pane user interface. The iPod has its own media-browsing interface on a 2.5-inch screen that can be operated using a touch-sensitive scroll-wheel and a few buttons. The two devices were designed to be used together, so much so that the iPod cannot be used effectively without iTunes. These two devices complement each others' functionality, interact and depend upon each other, and collectively fulfill the user's goal of listening to music or watching videos at the desk or when mobile.

In contrast, consider an example where the user's devices do not form an ecosystem: users work on documents on different devices, say, a work computer and a home computer, but when moving files between the two devices, they struggle to keep track of their files on both devices. In this process, they run the risk of overwriting a newer version with an older one and are forced to perform additional steps when trying to orchestrate this migration themselves. These two devices do not perform as a synergistic whole; they place on the user the burden of synchronization and version control.

Comparisons between technological and biological entities have been drawn in the past. For example, Nardi and O'Day define an information ecology as \textit{``a system of people, practices, values, and technologies in a particular local environment''} \cite{nardi_2000_information}. Their approach takes a more social view of the information ecology; ours focuses on a user's multiple devices and how the interaction among those devices influences the user's information management practices.

\subsection{Organisms}

In a biological ecosystem, there are two types of components: \textit{biotic} (living organisms), and \textit{abiotic} (environmental factors), both of which are equally important to the survival and development of the ecosystem as a whole \cite{tansley_1935_use}. While in the biological ecosystem the distinction is between living and non-living components, in our information ecosystem this demarcation is based on the component's ability to transform or transmit information.

\subsection{Information Flow Among Components}

In a natural ecosystem, energy flows from one component to another: chlorophyllous plants synthesize it from sunlight and other raw materials, while other biological entities receive their share from plants, directly or indirectly, forming the food chain.
In a device ecosystem, information can be considered the equivalent of energy. Some devices are producers of information: they capture or create information from the user (via input peripherals) or the environment (via sensors or external information sources such as the Web). It is then passed on to devices whose primary function is to process it. Certain other devices perform the role of consuming or disseminating information to satisfy the user's needs, via output devices or by exporting it to entities outside the ecosystem.

\subsection{Variety and Diversity}

One of the key characteristics of a biological ecosystem is the diversity of species found within. Similarly, a personal information ecosystem can contain a rich assortment of devices, differing in many ways (e.g. form factor, connectivity). 

The diversity of components in a personal information ecosystem is with respect to the capabilities of the devices to transform and transmit information. For example, large environmental displays can show status information (e.g. temperature, news), but do not allow any interaction. Laptop computers, however, also can show this information and, in addition, allow the user to create and transmit information. Some devices are hardly noticeable and blend in with the environment to stay invisible from the human eye (like micro-organisms). Described by Weiser in his vision of Ubiquitous Computing \cite{weiser_1991_the-computer}, these devices fulfill a very important role nevertheless. 

\subsection{Interdependencies}

Various species in a natural ecosystem depend on each other for various reasons: the flow of energy (or, analogously, information) among them is a key motivator. In biological systems, some of these relationships exhibit unique characteristics and are thus given special names, such as symbiosis, parasitism, and commensalism. We draw parallels between them and personal information ecosystems here.

\subsubsection{Symbiosis}
Symbiosis is defined as a relation between two kinds of organisms in which one obtains food or similar benefits from the other, while the latter benefits from this partnership. Similarly, two or more devices may offer complementary functionality, and depend upon each other to perform their task well. Each brings to the table a unique feature that is not found in the other, which is the \textit{raison d'\^etre} behind symbiosis. For example, PDAs offer the advantage of mobility, while desktops offer higher storage, processing power, and richer interaction paradigms. Using both devices symbiotically increases the value of the user's information by making it available from multiple places.

\subsubsection{Commensalism}
Commensalism is defined as a relation between two kinds of organisms in which one obtains food or other benefits from the other, but neither damaging nor benefiting it \cite{jaeger_1953_source-book}.
Likewise, as part of a device's natural function, it may provide, or broadcast, information that other devices use. For example, calendaring programs such as Apple iCal can publish a user's calendar for use by external entities. RSS feeds are routinely generated by web applications to disseminate information. Both of these practices help support other devices without hindering their own functionality; in the iCal example, the sharing of calendars is totally transparent to the user.

\subsubsection{Parasitism}
Parasitism represents a partnership in which one kind of organism obtains food or other benefit from another, and harms the host organism in the process. 
A notable example of such a relationship was mentioned by Paul Dourish during his keynote speech at the 2$^{nd}$ Latin American Conference on HCI, 2005: while at home, his Bluetooth-enabled phone headset located in his car would answer his cellphone when the cellphone was within the range of the Bluetooth headset. This behavior takes away the ability of the cellphone to be used on its own while in proximity of the headset, and thus we consider it a form of parasitism.
Thus, introduction of a new device into an ecosystem should not impede the existing relationships and balance in it.

\subsection{Environment}

In a biological ecosystem, various abiotic factors (physical as well as chemical) such as water, temperature, sunlight, etc. influence the organisms that live within it. They provide the infrastructure upon which life depends and thrives. Likewise, in a personal information ecosystem, factors such as the available power sources, network connectivity (either wired or wireless), cables and wiring, etc. help sustain the devices within. 
Similar to how fluctuations in the environment of an ecosystem affect its equilibrium, the changes in the support infrastructure of a personal information ecosystem can have far-reaching consequences for the devices within it.

In the natural world, there is a reciprocity between the environment and the species in it \cite{gibson_1977_theory}. Similarly, in the technology world, advances in the available infrastructure spur innovation in devices, which further drives infrastructure development, thus completing a full circle.

In some respects, organisms that depend on a critical environmental resource migrate towards a location where that resource is in abundance. Thus, we also see the adaptation of parts of a device ecosystem towards environments that can better sustain the system.
For example, datacenters are often established near sources of cheap power and near Internet backbone peering points. 
Users are attracted towards coffee shops that provide free wireless Internet connectivity.

\subsection{Processes}
The dynamism in a natural ecosystem comes from the various processes that occur naturally and continually in all organisms. Some of them are \textit{internal} to an organism, while some are \textit{between} two organisms. In an analogous fashion, continual internal processing and inter-device communication occurs in the devices in an ecosystem. 
For example, a paper being scanned into an image is an example of information interchange between the environment and a scanner device, while analyzing and interpreting the text via optical character recognition is an example of internal processing.

\subsection{Equilibrium}

A healthy ecology is not static, even when it is in equilibrium \cite{nardi_2000_information}. Subtle changes in the composition of an ecosystem or variations in the abiotic or biotic factors necessitate a response by the ecosystem as a whole to maintain equilibrium. However, at times, the introduction of certain species into an otherwise-balanced ecosystem, or the removal of certain critical species (either \textit{foundation species} or \textit{keystone species}) may impact the ecosystem adversely and cause it to lose its equilibrium.

Correspondingly, we define a personal information ecosystem to be in \textit{equilibrium} when the user's information needs are met and the information flow and interdependencies remain stable over a period of time. Components may be added to or removed from the ecosystem at any time.
Sometimes, the addition of a component leads to gradual \textit{evolution} of the information flow. 
Replacing a desktop computer with another one of higher processing power, but running the same applications on the same data as before, represents a low-impact change to the ecosystem since it does not affect the flow of information in any significant manner. In other cases, the introduction of an incompatible device type may completely disrupt the equilibrium. For example, the introduction of a PDA into an ecosystem may involve drastic changes to the information flow and management practices of the user.

As users progress from one stage of their life to another, their information needs and management strategies change. For example, a high school student probably maintains his/her calendar schedule only, whereas a college student also needs to be aware of the schedule and office hours of his/her professors and teaching assistants. Such changes in information needs often require additional devices or changes in workflow (e.g., a form of evolution) in the user's information ecosystem.

\section{Implications for Design}

Design practices have evolved over time. As McCullough \cite{mccullough_2004_digital} [p.152] eloquently puts it, ``Technology-centered interface design becomes human-centered interaction design. Emphasis on the solo task gives way to emphasis on social processes. Optimization for performance specifications or first-time usability metrics gives rise to whole-systems engineering for configurations we can live with, master, and tune.''

We argue that a new step in the evolution of design practices is needed to account for the synergistic use of multiple devices to fulfill a user's information needs. The parallels we drew between biological ecosystems and personal information ecosystems offer us a new way to think about the design space for multi-platform interfaces. Designers now have the responsibility of thinking about the equilibrium of the user's information ecosystem when designing new devices.

When designing for an ecosystem of devices, it is necessary to consider all platforms together and distribute or replicate functionality according to the affordances and contexts of use of each device \cite{pyla_2006_multiple}. This may require forfeiting interface-level consistency between two or more platforms in favor of presenting a `holistic' interface to the user. A holistic design approach would eliminate issues like in the alarm example described in the Introduction. 

Based on our experience, we have identified a number of issues to be considered when designing devices for a personal information ecosystem. The following sections present some of these issues briefly.

\subsection{Synchronization Is Not Enough}

Synchronization of data between two devices only means that their internal state will be made as consistent as possible. Simple synchronization implies nothing about the user's interaction with those devices. Synchronization is often error-prone and potentially introduces more file management tasks for the user to deal with. 
From prior research \cite{barreau_1995_finding}, we know that file management is a cognitively demanding task. Thus, if synchronization increases the file management tasks of a user then the user is subjected to higher cognitive load.

\subsection{Task Disconnects and Seamless Task Migration}

A \textit{task disconnect} represents the break in continuity that occurs due to the extra actions necessary when a user attempts to switch devices to accomplish a single task \cite{pyla_2006_multiple}.

Task disconnect is the cost of moving work from one device to another. Requiring the user to manage the information flow between devices interrupts the information flow in the ecology. We consider seamless task migration a principal attribute of an ecosystem in equilibrium. It is the designer's responsibility to support seamless task migration in a user's personal information ecosystem with the products they create. Seamless task migration is partially dependent on knowledge continuity and task continuity \cite{denis_2004_inter-usability}. 

\section{Conclusions}

In this paper, we presented the definition, characteristics and examples of a personal information ecosystem. We propose that such a view of personal information management and the plethora of devices in existence today is essential for the design of future ubiquitous environments. In our research, we have encountered numerous examples of devices that, when used individually, satisfy traditional usability requirements; however, when considered as part of a personal information ecosystem, they tend to disrupt the users' information flow and throw the ecosystem out of balance.

Future research should study how to evaluate the equilibrium of a personal information ecosystem and to assess the short- and long-term impact of the introduction and removal of devices. Furthermore, we should identify other important attributes of an ecosystem.

\section{Acknowledgments}

The authors would like to acknowledge the efforts of Lukena Karkhanis in reviewing a draft of this paper.


\newpage

\begin{thebibliography}{10}

\bibitem{barreau_1995_finding}
D.~Barreau and B.~A. Nardi.
\newblock Finding and reminding: file organization from the desktop.
\newblock {\em SIGCHI Bull.}, 27(3):39--43, 1995.

\bibitem{bergman_2006_the-project}
O.~Bergman, R.~Beyth-Marom, and R.~Nachmias.
\newblock The project fragmentation problem in personal information management.
\newblock In {\em CHI '06: Proceedings of the SIGCHI conference on Human
  Factors in computing systems}, pages 271--274, New York, NY, USA, 2006. ACM
  Press.

\bibitem{boardman_2004_stuff}
R.~Boardman and M.~A. Sasse.
\newblock "stuff goes into the computer and doesn't come out": a cross-tool
  study of personal information management.
\newblock In {\em CHI '04: Proceedings of the SIGCHI conference on Human
  factors in computing systems}, pages 583--590, New York, NY, USA, 2004. ACM
  Press.

\bibitem{denis_2004_inter-usability}
C.~Denis and L.~Karsenty.
\newblock Inter-usability of multi-device systems - a conceptual framework.
\newblock In A.~Seffah and H.~Javahery, editors, {\em Multiple User Interfaces:
  Cross-Platform Applications and Context-Aware Interfaces}, pages 373--384.
  John Wiley and Sons, 2004.

\bibitem{gibson_1977_theory}
J.~J. Gibson.
\newblock {The Theory of Affordances}.
\newblock {\em Perceiving, acting and knowing: toward an ecological
  psychology}, pages 67--82, 1977.

\bibitem{jaeger_1953_source-book}
E.~C. Jaeger and I.~H. Page.
\newblock {\em A source-book of medical terms.}
\newblock Charles C. Thomas, Springfield, Ill, 1953.

\bibitem{mccullough_2004_digital}
M.~McCullough.
\newblock {\em {Digital Ground: Architecture, Pervasive Computing, and
  Environmental Knowing}}.
\newblock MIT Press, 2004.

\bibitem{nardi_2000_information}
B.~Nardi and V.~O'Day.
\newblock {\em {Information Ecologies: Using Technology with Heart}}.
\newblock MIT Press, 2000.

\bibitem{perry_2001_dealing}
M.~Perry, K.~{O'Hara}, A.~Sellen, B.~Brown, and R.~Harper.
\newblock Dealing with mobility: understanding access anytime, anywhere.
\newblock {\em ACM Trans. Comput.-Hum. Interact.}, 8(4):323--347, 2001.

\bibitem{pyla_2006_multiple}
P.~S. Pyla, M.~Tungare, and M.~P\'{e}rez-Qui\~{n}ones.
\newblock {Multiple User Interfaces: Why Consistency is Not Everything, and
  Seamless Task Migration is Key.}
\newblock In {\em Proceedings of the CHI 2006 Workshop on The Many Faces of
  Consistency in Cross-Platform Design.}, 2006.

\bibitem{tansley_1935_use}
A.~G. Tansley.
\newblock The use and abuse of vegetational concepts and terms.
\newblock {\em Ecology}, 16(3):284--307, 1935.

\bibitem{weiser_1991_the-computer}
M.~Weiser.
\newblock The computer for the 21st century.
\newblock {\em Scientific American}, 265(3):66--75, 1991.

\end{thebibliography}
\end{document}